\documentclass[twoside]{article}
\usepackage{fleqn,espcrc2,epsfig}

\newcommand{\be}{\begin{equation}}
\newcommand{\ee}{\end{equation}}

\newcommand{\AmS}{{\protect\the\textfont2
  A\kern-.1667em\lower.5ex\hbox{M}\kern-.125emS}}

\title{Neutrino mass spectrum and lepton 
mixing
\thanks{Talk given at the XIX International Conference 
on Neutrino Physics and Astrophysics, Neutrino-2000,
Sudbury, Canada, 16 - 21 June 2000 (transparencies are available at
http://nu2000.sno.laurentian.ca). }}

\author{A. Yu. Smirnov \address{
International Center for Theoretical Physics, \\
Strada Costiera 11, 34100 Trieste, Italy }}

\begin{document}

\begin{abstract}

The program of 
reconstruction of the neutrino mass and flavor spectrum is 
outlined and the  present status of research is summarized. 
We describe the role  of future solar and atmospheric neutrino
experiments,  
detection of the Galactic supernovae and double beta decay searches 
in accomplishing this program. 
The LSND result and four neutrino mass spectra are considered in
connection
with recent searches for the sterile components in the solar  and
atmospheric neutrino fluxes.   
\end{abstract}

\maketitle

\noindent
{\bf 1. INTRODUCTION.}\\

\noindent
{\bf 1.1. Two remarks.} 

There is a hope that {\it detailed} information on the neutrino mass
spectrum and lepton mixing may eventually shed the  light  on\\ 
$\bullet$~the origin of the neutrino mass,\\ 
$\bullet$~quark-lepton symmetry, unification of quarks and leptons, Grand
Unification,\\
$\bullet$~fermion mass problem,\\
$\bullet$~physics beyond the standard model in general.\\ 
``Detailed information" are the key words:   just  knowledge 
that masses are small is not enough to clarify the points. 

Results  on atmospheric neutrinos~\cite{SK}  
show that  the simplest possibility --  
hierarchical mass   spectrum with  small flavor 
mixing has  not been realized. 
The guideline 
from the quark sector is lost. 
In this connection  we should 
consider without prejudice  all possible mass and mixing spectra 
which  do not contradict experiment. \\

\noindent
{\bf 1.2. The present status.} 

There are three leptonic flavors: 
$\nu_{\alpha}$, $\alpha = e, \mu, \tau$ and at least three 
neutrino mass  eigenstates 
$\nu_i$ with eigenvalues  $m_i$ ($i = 1, 2, 3$). 
The program of reconstruction of the spectrum 
consists of the determination of\\  
$\bullet$~number of mass eigenstates,\\ 
$\bullet$~masses $m_i$, \\
$\bullet$~distribution of the flavor in the mass eigenstates 
described by the mixing matrix  $U_{\alpha i}$,\\  
$\bullet$~complex phases of $U_{\alpha i}$ and $m_i$.

What is the present status? 

The atmospheric neutrino data provide us with the most reliable 
 information. With high confidence level we can say that the
data  
imply  the $\nu_{\mu}$ oscillations  with  maximal  or near
maximal depth. 
Moreover, the oscillations are driven by non-zero  
$\Delta m^2$. From this  interpretation we can infer that 

(i) There is at least one mass eigenstate with 
\be
m_{a} \geq \sqrt{|\Delta m^2_{atm}|} \sim (4 - 6)\cdot 10^{-2}~~ 
{\rm eV} .
\label{mass}
\ee

Further implications depend on assumptions about the number of mass
eigenstates and  the type of mass hierarchy. 
In the case of 3$\nu$-spectrum with  {\it normal} mass hierarchy 
(fig.1),  
$m_3 \gg m_2, m_1$,  the heaviest state $\nu_3$ has  the mass
(\ref{mass}). 
If  the spectrum has the {\it inverted} mass hierarchy, $\nu_3$ is the
lightest state, $m_3 \ll m_{a}$, 
and $\nu_{1}$, $\nu_{2}$ form a system of degenerate neutrinos 
with $m_2 \approx m_1 \approx m_{a}$. 
In the case of completely degenerate spectrum one has  
$m_3 \approx m_2 \approx m_1 \gg m_{a}$.  

(ii) The admixture of the $\nu_{\mu}$ flavor in the $\nu_3$ mass 
eigenstate is  
\be
|U_{\mu 3}|^2 = 0.3 - 0.7 ~~~~(90 \%~~{\rm CL}).
\label{ }   
\ee

(iii) The admixture of the electron neutrino  in the third  state is 
zero or small \cite{CHOOZ}:  
\be
|U_{e3}|^2 \leq  0.015 - 0.05. 
\ee
Thus, $\nu_{\mu}$ is mixed almost maximally
with  $\nu_{\tau}$ or/and $\nu_s$.

(iv) The $\nu_{\mu} - \nu_{\tau}$ channel gives better description of the
data than $\nu_{\mu} - \nu_s$ one: the latter is disfavored at 
3 $\sigma$ level \cite{STER}. Substantial contribution of the sterile 
channel is however possible already at  $2 \sigma$ level.

(v) We assume that $\nu_1$ and $\nu_2$ are responsible for the solution of
the solar neutrino problem. The best fit values of the 
oscillation parameters   
from all solution regions  (LMA, SMA, LOW, VO) satisfy inequality 
$$
\Delta m^2_{\odot} \ll \Delta m^2_{atm}.
$$
That is, the hierarchy of the $\Delta m^2$ exists. 

(vi) The distribution of the electron flavor depends on  
the solution of the $\nu_{\odot}$-problem.  

Clearly, with this information we are just in the beginning of realization
of the program.  
In what follows I will consider the next steps. \\

\begin{figure}[t]
\vspace*{-4.1truecm}
\hspace*{0.6truecm}
\epsfig{bbllx=1.4truecm,bblly=1.3truecm,bburx=19.5truecm,bbury=26.5truecm,%
height=8.0truecm,figure=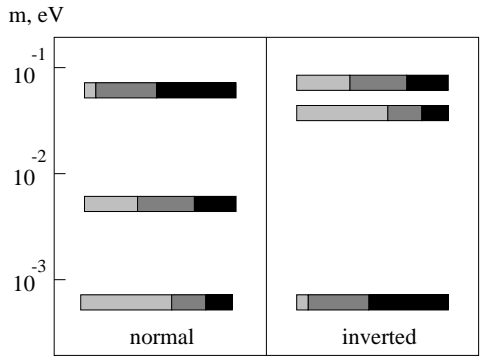}
\vspace*{-0.1cm}
\caption{The 3 $\nu$ mass spectra with normal and inverted mass
hierarchy. Boxes  show admixtures of different flavors in the mass
eigenstates: electron  (light grey), muon  (grey) and  
tau  (black).}
\label{f1}
\end{figure}

\noindent
{\bf 2. $U_{e3}$, HIERARCHY, DEGENERACY.} \\

\noindent  
{\bf 2.1. $U_{e3}$.}               

Future long-baseline experiments MINOS \cite{MINOS}, CERN-GS \cite{CG}
will be able 
to mildly improve present CHOOZ bound. An estimated sensitivity 
is at most  $|U_{e 3}|^2 \approx 5 \cdot 10^{-3}$ at 
$\Delta m^2_{atm} = 3 \cdot 10^{-3}$   eV$^2$. 
Further improvements of the reactor bound are rather difficult 
(see \cite{MIK}). Signatures of  non-zero $|U_{e 3}|^2$ exist in the 
atmospheric neutrinos. 
In fact,  best fit value
of $|U_{e 3}|^2$ from the atmospheric
neutrinos differs from zero, although the deviation is
statistically insignificant.
However, it is difficult to improve the situation with
present experiments and  the possibilities of future 
atmospheric neutrino detectors deserve special study.

Registration  of the neutrino bursts from the Galactic supernova 
by existing detectors SK, SNO  (several thousands events) will 
give  information about $|U_{e 3}|^2$ down to $10^{-5} - 10^{-4}$
\cite{DIGHE}. 

Even better  sensitivity $|U_{e 3}|^2 > 3\cdot 10^{-5}$ may be achieved at
the neutrino factories \cite{NUFAC}. 

Intuitively, it is difficult to expect very small $|U_{e 3}|^2$ 
if mixing between the second  and the third generation is almost maximal 
and the mixing of the electron neutrinos is also maximal or large   
(unless some special arrangements are done).
This has been quantified recently in terms of the neutrino mass matrices
which lead to the 
solutions of the solar and atmospheric neutrino problems \cite{AKH}.  
In the assumption that there is no  special fine tuning 
of  the matrix elements  $m_{e \mu}$ and $m_{e \tau}$,  so that 
$|m_{e \mu} - m_{e \tau}| \sim max [m_{e \mu}, m_{e \tau}] $, the
following 
relation has been found \cite{AKH}: 
\be
U_{e3}^2 \approx \frac{1}{4} 
\frac{\tan^2 2\theta_{\odot}}{\sqrt{1 + \tan^2 2\theta_{\odot}}}
\frac{\Delta m^2_{\odot}}{\Delta m^2_{atm}}. 
\label{relation}
\ee
For parameters from the LMA region we get values $|U_{\mu 3}|^2 = 0.003 -
0.02$, where the upper edge is the 
the present experimental bound. \\

\noindent  
{\bf 2.2. Hierarchy and Degeneracy.}               

Phenomenology of  schemes with normal and inverted
mass hierarchy is different. The hierarchy can be identified by studies 
of (i) the neutrinoless double beta decay, (ii) Earth matter effect on 1 -
3 mixing in the atmospheric neutrinos and in the long baseline
experiments, (iii) neutrino burst from supernovae. 
 
In the scheme with inverted hierarchy the  contribution of neutrinos  to
the energy density of the Universe can be two times larger than that in
the scheme with normal  hierarchy: 
$\Omega_{\nu} \geq 2 \sqrt{\Delta m^2_{atm}} n_{\nu}$   
( $n_{\nu}$ is the concentration of one neutrino species).  
The scheme with normal mass hierarchy
(or in general,  the
scheme with $\nu_3$ being the heaviest state)
may have partial degeneracy when $m_2 - m_1 \ll m_2$.
In this case for a given oscillation pattern both
$\Omega_{\nu}$ and $m_{ee}$ can be larger than in the
hierarchical case.\\

\noindent  
{\bf 3. IDENTIFYING SOLUTION OF THE $\nu_{\odot}$-PROBLEM.} \\

Identification of the solution is one of the major steps in the
reconstruction of the spectrum which will significantly determine
further strategy of the research. It will allow us to:  
(i) measure $\Delta m^2_{21} \equiv \Delta m^2_{\odot}$ as well as the  
distribution of the electron flavor: $|U_{e1}|^2$,  $|U_{e2}|^2$;  
(ii) find or restrict the presence of the sterile neutrinos,  
(iii) estimate a possibility to  measure  the CP-violation  
and to  discover the neutrinoless beta decay. 

Let us describe some recent results. \\

\noindent  
{\bf 2.1. Flux during the night.} 

The zenith angle  ($\theta_Z$) distributions  of events during the  nights
differ  for different solutions and therefore 
precise measurements  of the distribution can be used to discriminate
among the solutions.

The LMA solution predicts rather flat distribution of events 
with slightly lower rate in the first night bin N1 
($\cos \theta_Z = 0 - 0.2$). 
The reason is that the oscillation length is small 
and substantial averaging of oscillations occurs in 
all the bins \cite{LMA}. 

For the LOW solution, the  maximal rate is expected in the second night
bin N2 ($\cos \theta_Z = 0.2 - 0.4$) \cite{LOW}. Indeed, 
for  parameters from the LOW region, the oscillation length 
in matter is determined basically by the refraction length,  
$l_m \approx l_0$,  and it depends weakly 
on  $E/\Delta m^2$ and mixing. 
No averaging occurs. It turns out that the average length of the 
neutrino trajectories in the N2 bin equals half of the refraction length, 
so that the 
oscillation effect is maximal. 
The length of the trajectory is about $l_0$ in the N3 bin, where
minimum of the rate is expected. 
The height of the peak in N2 bin  decreases with $\Delta m^2$.

In the case of SMA solution maximal rate  is expected in the N5 (core) 
bin \cite{SMA}, where the parametric enhancement of oscillations can take
place. The peak decreases with mixing angle and at  
$\sin^2 2\theta \sim 3 \cdot 10^{-3}$ it  transforms to the 
deep. 

No enhancement of the night rate should be seen for VO solution. 

Thus, using information on integrated day-night asymmetry and  signals in 
N2 and N5 bins one can identify the solution.

Notice that the zenith distribution observed by SK 
does not fit  any of expected  distributions: maximal rate is in the N1 
bin, and there is no enhancement of rate neither in N2 nor in N5 bins. 
In the SNO the expected zenith angle distributions have similar character, 
however absolute value of the regeneration effect is 
larger due to absence of  damping related  to $\nu_{\mu}$ and
$\nu_{\tau}$   contribution to  the SK signal. \\

\noindent  
{\bf 3.2. Correlations of observables.} 

Present searches for the ``smoking guns" of  
certain solutions of the $\nu_{\odot}$-problem give just 
$(1 - 2)\sigma $ indications. 
To enhance the identification
power of the analysis we suggest  to study correlations of 
various observables \cite{CORR}. 
Indeed, correlations of  observables appear  for 
different  solution of the $\nu_{\odot}$-problem  
and they  can be considered as signatures of corresponding 
solutions.

\begin{figure}[t]
\vspace*{-0.8truecm}
\hspace*{0.1truecm}
\epsfig{bbllx=1.4truecm,bblly=1.3truecm,bburx=19.5truecm,bbury=26.5truecm,%
height=9.6truecm,figure=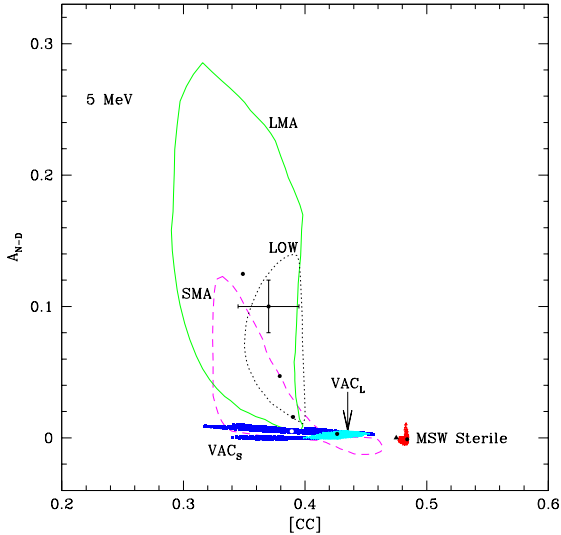}
\vspace*{-2.5cm}
\caption{The allowed regions for the day-night asymmetry versus 
reduced rate (CC events at SNO above 5 MeV). 
The best
fit points for each solution are indicated by black circles and
no-oscillation case is shown by a triangle. 
The cross is a simulated measurement with  1 $\sigma$ 
error bars. (From  
\protect\cite{CORR}).}
\label{f4}
\end{figure}

The observables (denote them by $X$, $Y$) include rates of events 
at different detectors, characteristics of spectrum distortion 
(e.g., shift of the first moment of the spectrum) 
and parameters of the time variations of signals 
(day-night asymmetry, seasonal asymmetry, etc.).  
To find the correlations we have performed the mapping of the  solution
regions in the $\Delta m^2$ - $\sin^2 2\theta$ plane onto the plane  of
observables $X$ and $Y$. If $\Delta m^2 - \sin^2 2\theta$ region 
projects onto the line in the $X - Y$ the correlations is  
very strong. In general, the criteria for strong correlation is that the
area of the projected region, $S_{XY}$, is much smaller than  the product 
$\Delta X \times \Delta Y$, where $\Delta X$ and  $\Delta Y$ are 
allowed intervals of $X$ and $Y$  when they are treated independently. 

In fig.2  we show, as an example, mapping of the 
$\Delta m^2 - \sin^2 2\theta$ 
regions of solutions onto the plane of the SNO observables [CC] and 
$A_{DN}$, where [CC] $\equiv N_{obs}/N_{SSM}$  is the reduced rate of the  
charged current events  and $A_{DN} \equiv 2(N - D)/(N + D)$ is the 
day-night asymmetry of the charged current events. \\

\noindent  
{\bf 3.3. Large or Maximal?}

The three among five solutions of the solar neutrino problem require 
large mixing angle of the electron neutrino. Moreover,  the LMA 
solution gives the best global fit of the data.  
The best fit of the atmospheric neutrino data corresponds to maximal
mixing. 
Is large  (or maximal) mixing the  generic property of leptons? 
What is a deviation from maximal mixing? 

These questions are 
important for theory \cite{NIR}: The deviation from maximal 
mixing can be related to small parameter $\lambda \sim 0.22$ which 
characterizes the fermion mass hierarchy and appears in the theories with
flavor symmetry. We describe the deviation by 
$\epsilon \equiv \cos 2 \theta$, 
($\epsilon = 0$ at the maximal mixing). 
Depending on  model one can get $\epsilon = \lambda^n$, where 
usually $n = 1$ or  $2$, or  
$\epsilon \sim \sqrt{m_e/ m_{\mu}} = 0.07$, or   $\epsilon \sim m_e/
m_{\mu} = 0.005$, etc.. 

In contrast with theory,   maximal mixing is not a special point for 
phenomenology. Nothing dramatic happens when 
$\epsilon$ changes the sign: no divergencies or discontinuities 
appear, all observables depend on $\epsilon$ 
rather smoothly \cite{smooth,LOW}.  

Performing a global fit of all available solar neutrino data 
we find \cite{LOW} that maximal mixing is allowed  in the LMA region 
at 99.9 \% CL, and  in LOW region at 99 \% CL. For 
$\epsilon = 0.07$ the interval $\Delta m^2 = (2 - 30) \cdot 10^{-5}$ 
eV$^2$ (LMA)  is accepted at 99 \% CL,  etc.. 

Future perspectives of measurement of the deviation depend
significantly on the range of $\Delta m^2$. 
Observables depend linearly on $\epsilon$  in 
the range of the  MSW conversion (LMA,  LOW). 
In particular, the survival probability is proportional to 
$(1 - \epsilon)$, the day-night asymmetry  $\propto (1 + \epsilon)$, 
the distortion of spectrum 
$\propto  \epsilon$. 
In contrast, in the VO regions the dependence of observables 
on $\epsilon$ is quadratic: $\propto (1 + \epsilon^2)$ in the average
oscillation case and 
$\propto (1 - \epsilon^2)$ in the non-averaged case. 

Thus, for small $\epsilon$ the sensitivity of  measurements of the
deviation is much higher in the MSW regions of $\Delta m^2$. 
The most precise determination of $\epsilon$ will be possible with 
the SNO results. Simultaneous measurements of the double ratio  
[NC]/[CC] (of the neutral to  charged current reduced rates) and the
day-night asymmetry of the 
CC events will allow to determine $\epsilon$ with accuracy 
$\Delta \epsilon \sim 0.07$ ($1 \sigma$) \cite{LOW}. \\

\noindent
{\bf 4. MORE INFORMATION.}\\ 

\noindent  
{\bf 4.1.  Double beta decay  and test equalities.}

Remarks: 

  For any oscillation pattern (values of $\Delta m^2$ and 
$|U_{\alpha i}|^2$) the 
effective Majorana mass of the electron neutrino 
relevant for the $\beta \beta_{0\nu}$ decay,  
$m_{ee}$,  can take any value  from zero to experimental
upper bound  for the normal mass hierarchy and $|m_{ee}|$ has non-zero 
minimum value for the inverted hierarchy provided that $|U_{e 1}|^2 >
1/2$ \cite{DOUBLE}.

 If the neutrinoless double beta decay 
will be discovered and the rate
will give $m_{ee}$, then under assumption 
that the Majorana masses are the
only source of the decay, 
we can say that at least one mass should satisfy 
inequality $m_j > m_{ee}/n$, where $n$ is the number of mass 
eigenstates \cite{DOUBLE}.

 Definite predictions for $m_{ee}$ can be given  in the context of
certain 
neutrino mass spectra 
(see  \cite{PET,DOUBLE} and references therein):  $m_{ee}$ can be 
related (especially in the cases when dominant contribution 
comes from one mass eigenstate) with the
oscillation parameters. 
Therefore coincidence of the measured value $m_{ee}$ with some 
combination of the oscillation parameters will testify for certain 
neutrino mass spectrum. We can call these relations the 
{\it test equalities}. 

Let us give  examples of the test equalities: 

1). In the case of normal mass hierarchy, and SMA, LOW or VO solutions 
of the solar neutrino problem the dominant contribution comes 
from $\nu_3$: 
$$
m_{ee} \approx \sqrt{\Delta m^2_{atm}}|U_{\mu 3}|^2. 
$$

2). For normal mass hierarchy, LMA solution  and small 
$|U_{e 3}|^2$ one gets 
\be
m_{ee} \approx   \sin^2 \theta_{\odot}  \sqrt{\Delta m^2_{\odot}} 
\leq 5 \cdot 10^{-3}{\rm eV}
\label{norml-bb} 
\ee
which can be tested by  the 10 ton version of  GENIUS experiment.  	

3). In the case of inverted mass hierarchy, SMA  or 
LMA solutions with  equal phases of the mass eigenvalues 
$m_1$ and $m_2$,  one gets \cite{PET,DOUBLE}
\be
m_{ee} \approx  \sqrt{\Delta m^2_{atm}} = (5 - 7)\cdot 10^{-2} {\rm eV} 
\label{invl-bb}
\ee
which can be achieved already in the next generation of the double beta
decay experiments.    

4). For  inverted mass hierarchy and large mixing   
solutions with relative phase of the degenerate states  
$\phi_1 - \phi_2 = \pi$ (which holds for the pseudo Dirac system)
we find  
$$
m_{ee} \approx  \cos 2\theta_{\odot} \sqrt{\Delta m^2_{atm}}.  
$$ 

As follows from the above analysis, 
the bound 
$m_{ee} < 3 \cdot 10^{-3}$ eV will testify for the normal hierarchy, 
whereas values $m_{ee} > 10^{-2}$ eV are  the signature of the inverted
mass hierarchy provided that $\sin^2 2\theta_{\odot} \leq 0.9$. 
Predictions overlap in the case of partial or complete
degeneracy of spectra.\\

\noindent
{\bf 4.2. Supernova neutrinos.} 

In the three neutrino scheme there are two relevant resonances: 
high (density) resonance at $\rho_h \sim 10^{4}$ 
g/cm$^3$ 
the  low resonance at $\rho_l \sim 10 - 100$ g/cm$^3$
related to $\Delta m^2_{atm}$ and $\Delta m^2_{\odot}$ correspondingly. 
Since at the production point $\rho \gg \rho_h, \rho_l$, the supernova
neutrinos probe whole neutrino mass spectrum. 
Moreover, mixings
associated with both $\Delta m^2$ can be matter enhanced.

In spite of uncertainties related to  the density profile and,  
especially, to parameters 
of the original spectra, 
some  observables are largely {\it supernova model independent}  
which opens the possibility to get reliable
information on neutrino mass spectrum. In particular, 
inequalities of the average energies: 
\be
E(\nu_e) <  E(\bar{\nu}_e) < E(\nu_{\mu})
\label{ineq}
\ee
are SN model independent.  
Violation of these inequalities will testify for the  neutrino conversion. 
It is expected that spectra emitted during short time intervals 
$\Delta t \ll 10$ s are ``pinched". Observation of  wide spectra 
will testify for its compositness which appears as a result
of conversion.

Another SN model independent possibility is to study the Earth matter
effects on the 
neutrino fluxes from supernovae. The oscillations of the 
SN neutrinos in the matter of the Earth  can induce  irregular
structures in the otherwise  smooth energy spectra. These oscillations 
will lead also to  different  signals at different detectors   
(for which neutrino trajectories in the Earth are different). 

The observable effects are the result of the interplay of
neutrino conversion inside the star and oscillations inside the Earth. 
The level crossing schemes are different for normal and inverted mass
hierarchies. The high resonance is in the 
neutrino channel if the hierarchy is normal and it is in the antineutrino 
channel for the inverted mass  hierarchy. This can lead to completely
different 
patterns of conversion.

If $|U_{e 3}|^2 > 10^{-3}$, the conversion in the high resonance is
completely adiabatic. Taking into account also that original 
$\nu_{\mu}$ and $\nu_{\tau}$ fluxes are practically identical one gets
that in the normal hierarchy case: 
(i) $\nu_e$ converts completely to 
$\nu_{\mu} / \nu_{\tau}$, 
(ii) at the Earth $\nu_e$ should have hard
spectrum of the original $\nu_{\mu}$ and (iii) the Earth matter effect
does not influence  this flux. In contrast, the oscillation effect in the
matter of the Earth can be observed in the $\bar{\nu}_e$-spectrum. 

In the case of  inverted mass hierarchy $\nu_e$ and $\bar{\nu}_e$ 
interchange the roles: (i) $\bar{\nu}_e$  transforms in the star 
into $\bar{\nu}_{\mu} / \bar{\nu}_{\tau}$, (ii)  at the surface of the
Earth 
$\bar{\nu}_e$-flux will have a hard spectrum, 
(iii) the Earth matter effect
should not be seen in the $\bar{\nu}_e$ signal but it can be observed in
the $\nu_e$-signal. 

Thus,  the fact of observation of the Earth matter effect in 
the $\bar{\nu}_e$ flux, but not in $\nu_e$,  will testify for the normal
hierarchy. An opposite situation: the Earth matter effect in $\nu_e$ 
channel and an absence of the effect in $\bar{\nu}_e$ 
will be an evidence of the 
inverted mass hierarchy. Substantial matter effect 
is possible for  the LMA  parameters only. 

If $|U_{e 3}|^2 \ll 10^{-3}$ the high resonance is inefficient and
significant matter effect can be observed both in $\bar{\nu}_e$ and 
$\nu_e$ spectra. \\

\noindent
{\bf 5. $\nu_{\odot}$, $\nu_{atm}$ and LSND.}\\ 

\begin{figure}[t]
\vspace*{-3.1truecm}
\hspace*{1.5truecm}
\epsfig{bbllx=1.4truecm,bblly=1.3truecm,bburx=19.5truecm,bbury=26.5truecm,%
height=8.0truecm,figure=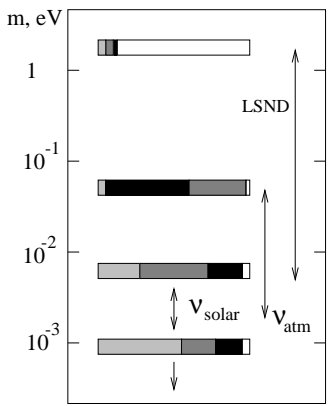}
\vspace*{-0.4cm}
\caption{The (3 + 1) spectrum of the neutrino mass. 
Boxes show the admixtures of the flavors in the mass eigenstates: 
electron  (light grey) muon (grey) tau (black) and sterile (white).   
}
\label{f4}
\end{figure}

It is widely accepted that simultaneous explanation of the solar,
atmospheric and LSND \cite{LSND} 
results in terms of oscillations requires an existence of the
sterile neutrino (see e.g. \cite{BIL}). 
Less appreciated fact is that the explanation requires  
the sterile neutrino to be a dominant component  in oscillations of solar
or 
atmospheric neutrinos. That is, either $\nu_e \rightarrow \nu_s$ is 
the dominant channel of the  solar neutrino conversion, or 
$\nu_{\mu}  \rightarrow \nu_s$ is the dominant oscillation mode 
for the atmospheric neutrinos. The extreme  situation is when 
sterile channels contribute 1/2  both in the solar and the atmospheric
neutrino transformations.

\begin{figure}[t]
\vspace*{0.4truecm}
\hspace*{-0.7truecm}
\epsfig{bbllx=1.4truecm,bblly=1.3truecm,bburx=19.5truecm,bbury=26.5truecm,%
height=10.6truecm,figure=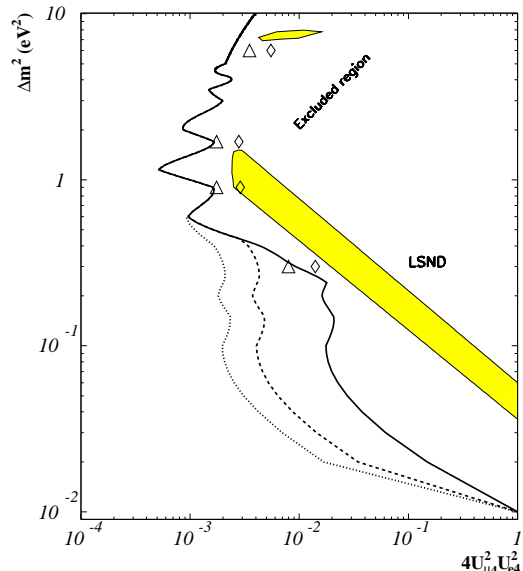}
\vspace*{-4.6cm}
\caption{The bounds on the effective mixing 
parameter $\sin^2 2\theta_{LSND}$ for the 
LSND experiment. Shown are: the product of the 
90 \% CL bound from BUGEY and CDHS experiments (solid line) 
Solid,  dashed and dotted lines 
below $\Delta m^2 \sim 0.2 - 0.3$ 
correspond to the  bound from the
atmospheric neutrinos for $\sin^2 2\theta_{23} = 1.0,~ 0.9,~ 0.8$ 
correspondingly. Rombs
and triangles show 99 \% and 
95\% CL bounds obtained from the BUGEY and CDHS results  
\protect\cite{ORLANDO}.}
\label{f4}
\end{figure}

This statement holds for the so called (2 + 2) 
scheme of the neutrino mass in which two pairs of the mass eigenstates
with mass splitting $\Delta m^2_{\odot}$ and $\Delta m^2_{atm}$ are
separated by the mass gap related to  $\Delta m^2_{LSND}$. In these  
schemes one easily gets the depth of 
$\bar{\nu}_{\mu} - \bar{\nu}_e$ oscillations required by the LSND. 
It is claimed \cite{BIL} that the (2 + 2) scheme is the only possibility
and the
alternative schemes give too small mixing for the LSND. Situation, 
however, may change: 

1. Recent  atmospheric neutrino data  disfavor  
$\nu_{\mu}  \rightarrow \nu_s$ as a dominant mode of oscillations. 
Although up to  0.5 contribution of the sterile
channel still gives a good fit \cite{SATM}. 

2. The mode   $\nu_e \rightarrow \nu_s$ although accepted, does not give
the best fit of the solar neutrino  data.  The  $\nu_e \rightarrow \nu_s$  
solution can be identified soon by 
(i) equality of the reduced charged current rate at SNO and 
electron scattering rate at SK: 
[CC] $\approx R_{SK}$;  (ii) small Day-Night asymmetry 
($< 2 \%$) in SK and SNO; (iii) unchanged double ratio  
[NC]/[CC]$ \approx 1$ (see \cite{CORR}). 

If it will be proven that both in the solar and atmospheric
neutrinos the contribution of the sterile component  is smaller than 1/2, 
the (2 + 2) scheme  should be rejected, and  the oscillation
interpretation of the LSND will be questioned.

In this connection we have reconsidered the (3 + 1) scheme 
(see fig.3) in which three mass eigenstates with
splittings $\Delta m^2_{atm}$ and $\Delta m^2_{\odot}$ form the flavor
block with small admixtures of sterile neutrino and the fourth state
(predominantly  sterile) 
is isolated from the flavor block  by the mass gap 
$\Delta m^2_{LSND}$  \cite{ORLANDO}. Both solar and
atmospheric neutrinos transform into active ones.  

The effective mixing parameter for 
$\bar{\nu}_{\mu} - \bar{\nu}_e$ oscillations driven by $\Delta m^2_{LSND}$ 
equals 
\be
\sin^2 2\theta_{LSND} = 4 U_{e4}^2 U_{\mu 4}^2, 
\label{LSND}
\ee
where  $U_{e4}^2$ and  $U_{\mu 4}^2$ are 
the admixtures of the $\nu_e$ and 
$\nu_{\mu}$ in the fourth mass eigenstate.  
$U_{e4}^2$ and  $U_{\mu 4}^2$ determine the 
$\nu_e-$ and $\nu_{\mu}-$ disappearance in oscillations driven 
by $\Delta m^2_{LSND}$ and are restricted by the results of  BUGEY
\cite{BUGEY} and CDHS \cite{CDHS} experiments correspondingly. 
For low values of $\Delta m^2$ better limit on $U_{\mu 4}^2$ follows 
from the atmospheric neutrinos \cite{BIL}. In fig.4  we reproduce the
bound on 
$\sin^2 2\theta_{LSND}$ obtained in \cite{BIL} using  Eq. (\ref{LSND}) and
the 90 \% CL bounds from BUGEY and CDHS. The bound 
excludes whole allowed LSND region which led to conclusion that
(3 + 1) scheme can not reproduce the LSND result. 
However, the question is:  which confidence level should be prescribed
to this bound?  

For several values of $\Delta m^2$ we have found 
the  95 \% CL and   99 \% CL bounds on 
$\sin^2 2\theta_{LSND}$ in assumption that distributions of 
the $U_{\alpha 4}^2$ ($\alpha = e, \mu$)  implied by the experiments are 
Gaussian (see fig. 4). We used central values 
of $U_{\alpha 4}^2$ and and 90 \% CL bounds 
published in the papers \cite{BUGEY,CDHS} to restore 
parameters of the Gaussian distributions. 

As follows from the fig.4,   in the range $\Delta m^2 \sim 1$ eV$^2$, 
the product of the 90\% CL bounds corresponds to $\sim 95 \%$  CL. 
The CL decreases with increase of $\Delta m^2$. At 99\% 
CL the  LSND region  at $\Delta m^2 \sim 1$ eV$^2$ 
becomes acceptable. Moreover, new analysis \cite{LSND00} 
shifts the allowed LSND region to  smaller $\sin^2 2\theta$, so that now 
some part of the region is acceptable even at 95\% CL. 

The (3 + 1) scheme leads to a number of the phenomenological   
consequences which can be checked in the forthcoming experiments. 
It has also interesting astrophysical and cosmological consequences 
\cite{ORLANDO}. \\

\noindent
{\bf 6. CONCLUSIONS.}\\ 

What are perspectives of the reconstruction of the neutrino mass and
flavor spectrum? 

1. Identification of the dominant mode of the atmospheric neutrino 
oscillations has a good chance with further studies at SK 
and LBL experiments. The bound on the presence of sterile
neutrinos will be better than $|U_{s3}|^2 < 1/2$, which  has 
important implications for  theory.  

2. Distribution of the electron flavor ($|U_{e1}|^2$, $|U_{e2}|^2$ 
as well as  $\Delta m^2_{\odot}$ will be  determined together with
identification of  the solution of the $\nu_{\odot}$-problem. 
Sooner or later (depending on our luck) this will be done by future 
measurement at SK, SNO, GNO, SAGE,  BOREXINO. Studies of correlations 
of observables will allow us to  enhance  an  identification power of  
analysis. Ironically, the solution of the $\nu_{\odot}$ problem can be
found  without solar neutrinos -- in KAMLAND experiment. 
Important bound will be obtained on presence of $\nu_s$.

There is some chance to measure $|U_{e 3}|^2$ in the 
forthcoming LBL experiments.

3. Determination of the type of mass hierarchy, the  
level of spectrum degeneracy, the 
CP-violating phase and the absolute scale on the neutrino mass
will require much more serious efforts. Progress will be related 
to the oscillation experiments  with very long base lines and 
probably with direct measurements of the
neutrino mass. 
Identification of the  type of  mass hierarchy and important 
bounds on value $|U_{e 3}|^2$ can be obtained  
by  the detection  of
the neutrino burst from the Galactic supernova. 
Observation of the $\beta \beta_{0\nu}$-decay means   
the  discovery of the lepton number violation and the Majorana 
nature of neutrinos. Measurements of the effective mass $m_{ee}$ 
will allow to check ``test equalities" which relate $m_{ee}$ and 
oscillation parameters in the context of certain schemes of neutrino mass.  
In this way we will be probably  able to identify the scheme 
and to get information on the absolute scale of neutrino mass.


\end{document}